\documentclass{aastex}
\usepackage{spr-astr-addons}
\usepackage{url}

\RequirePackage{color}

\begin{document}

\title{RAT0455+1305: another pulsating hybrid sdB star}
\slugcomment{}
%% Running heads
\shorttitle{RAT0455+1305}
\shortauthors{A.Baran \& L.Fox-Machado}

\author{Andrzej Baran\altaffilmark{1}}
\affil{Cracow Pedagogical University}
\and
\author{Lester Fox-Machado\altaffilmark{2}}
\affil{Instituto de Astronom\'{\i}a-UNAM}
%\email{\emaila}

\altaffiltext{1}{Iowa State University}
%\altaffiltext{2}{Second Alternate Affilation.}
%\altaffiltext{3}{Third Alternate Affilation.}

\begin{abstract}
RAT0455+1305 was discovered during the Rapid Temporal Survey which aims in finding any variability on timescales of a few minutes to several hours. The star was found to be another sdBV star with one high amplitude mode and relatively long period. These features along with estimation of T$_{\rm eff}$ and $\log g$ makes this star very similar to Balloon\,090100001. Encouraged by prominent results obtained for the latter star we have decided to perform white light photometry on RAT0455+1305. In 2009 we used the 1.5m telescope located in San Pedro Martir Observatory in Mexico. Fourier analysis confirmed the dominant mode found in the discovery data, uncovered another peak close to the dominant one, and three peaks in the low frequency region. This shows that RAT0455+1305 is another hybrid sdBV star pulsating in both p\,-- and g\,--\,modes.
\end{abstract}


%\section*{}
%\label{sec:intro}

\section{A short story of hybrid sdB stars}

Pulsations in hot subdwarf stars were discovered more than a decade ago \citep{kilk97}. After this accidental discovery there was a boom in searching for other hot subdwarf stars showing stellar oscillations. To date more than 50 objects of this kind have been found. Most of them show pressure modes (p\,--\,modes) only and are called V361\,Hya\,--\,type stars after the prototype. Those that pulsate in gravity modes (g\,--\,modes) are called V477\,Her\,--\,type. Unexpectedly some of the stars previously counted to the former group pulsate in both p\,-- and g\,--\,modes. The first member of the hybrid  group, DW\,Lyn (HS0702+6043) was found and published by \cite{schuh06}. A new refined time series re\,--\,analysis revealed, beside p\,--\,modes detected in the discovery data, at least one mode in the low frequency region which has been assigned as a gravity mode region with the periods of about ten times longer than those for the p\--\,modes. As we know that p\,-- and g\,--\,modes are rooted in different layers in the star, the presence of two kinds in one star is a plus for theoretical models and it allows to probe wider part of the star. It is obvious, however, that comparison of the frequency spectrum derived from observations with the theoretical one is more reliable if more modes are driven in a star. In case of DW\,Lyn, one g\,--\,mode is not promising for easy modelling. Only recently \cite{lutz08} found two other g\,--\,modes in this star. Another hybrid star discovered about the same time is Balloon\, 090100001 (Bal09). Analysis of data by \cite{baran05} shows, unlike to the discovery data, the hybrid nature of the star. Discovery of the next hybrid star was not as unusual as the number of modes in the low frequency domain. Almost ten g\,--\,modes make this object the most promising candidate for studying the interior of hot subdwarfs by means of asteroseismology. The latest hybrid star, V391\,Peg was found by \cite{lutz08}. The data on this star show only one g\,--\,mode.

\begin{figure}[]
\plotone{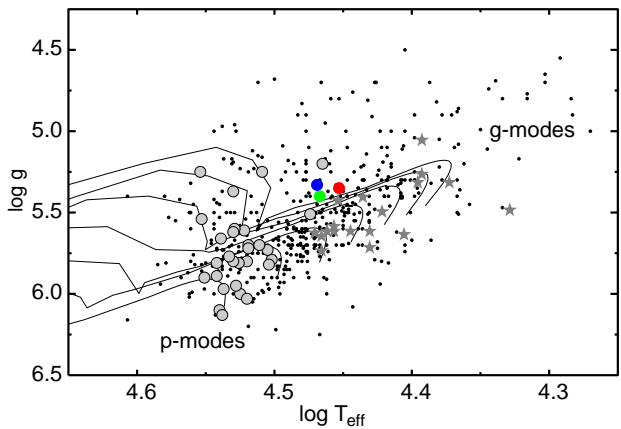}
\caption{$\log$T$_{\rm eff}$ and $\log g$ plane for sdB stars. See text for details} %% no full stop at the end
\label{fig1}
\end{figure}

\section{Where are the hybrids placed?}
Fig.\ref{fig1} shows the region of $\log$T$_{\rm eff}$ and $\log g$ where hot subdwarf stars are located. Evolutionary tracks from the ZAEHB for models with different mass of the hydrogen envelopes are plotted with solid lines. The gray circles indicate stars found to pulsate in pressure modes only while gray asterisks indicate objects with gravity modes detected. These groups are located in two different regions. Stars with p\,--\,modes are, in general, hotter and have stronger surface gravity whereas g\,--\,modes pulsators are cooler and have weaker surface gravity. There are also objects which seems to be right between those two groups. Three of them marked with color circles are the hybrid stars: blue\,--\,Bal09, red\,--\,DW\,Lyn, green\,--\,V391\,Peg. Their location, not predicted before discovery of the hybrid prototype, is nothing unusual. The most likely reason is that the condition inside the star is appropriate to drive both kinds of pulsations but at the different depth. But why the other stars, denoted by three gray circles and asterisks, placed almost in the same region as hybrid stars were not found to be hybrids? The newest data on RAT0455+1305 (RAT0455) confirms that this star is not an exception and it also has a hybrid nature.
As for the rest of the stars, perhaps more data are needed to confirm or exclude their hybrid nature. We postpone any further discussion about the hybrid nature of the other stars to the discussion section.

\begin{figure}[]
%\plotone{fig2.ps}
\includegraphics[angle=-90,scale=0.3,width=\columnwidth]{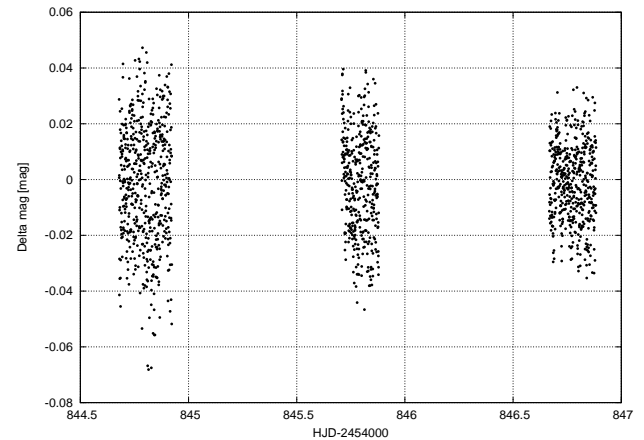}
\caption{Light curve in white light of RAT0455} %% no full stop at the end
\label{fig2}
\end{figure}

\begin{table}[t]
\caption{Observational log.}
\begin{tabular}{@{}cccc}
\hline
Date      & hours & exposure & filter \\
13 Jan 2009 & 5.8   & 30s     & none \\
14 Jan 2009 & 4.1   & 30s     & none \\
15 Jan 2009 & 5.2   & 30s     & none \\
\hline
\end{tabular}
\label{tab1}
\end{table}

\section{Discovery data}
RAT0455 has been classified as a pulsating hot subdwarf in the course of the RApid Temporal Survey (RATS) \citep{ramsay06}, which aims in finding any variability on timescales of a few mins to several hours. The survey reaches 22\,mag stars, so it includes objects which have not been observed or classified so far. The nature of many objects in this survey is still unknown. The main strategy in the survey is to take time\,--\,series photometry and spectroscopy only for objects suspected of variability on the basis of short time data. One of the star (RAT0455), which has been found to show variability, turned out to be a hot subdwarf. Fourier analysis of the data taken during the survey revealed one mode with period of about 6.23\,min and amplitude 40\,mmag (averaged from three nights). The brightness of this star is not impressive; it is only 17.2\,mag in V filter. For this reason, the spectral classification was not done so the star could not be included in the list of sample stars for photometric follow\,--\,up. Furthermore, even after this star classified as a pulsating hot subdwarf it did not raise quick interest in re\,--\,observation to discover more modes (if any).

\begin{figure}[]
%\plotone{fig3.ps}
\includegraphics[angle=-90,scale=0.3,width=\columnwidth]{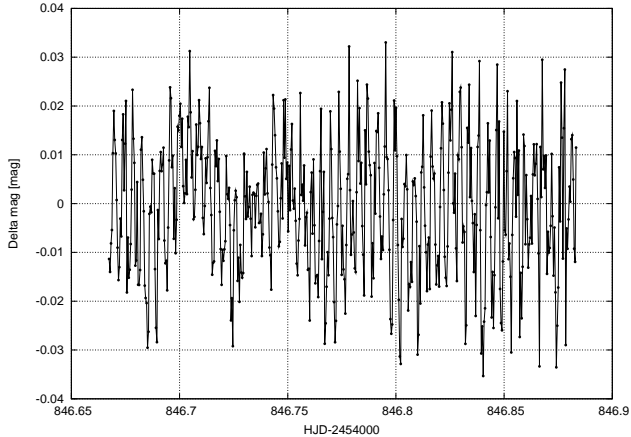}
\caption{One night of data on RAT0455. Beside short period oscillation a long one can also be noticed} %% no full stop at the end
\label{fig3}
\end{figure}

\begin{figure}[]
%\plotone{fig4.ps}
\includegraphics[angle=-90,scale=0.3,width=\columnwidth]{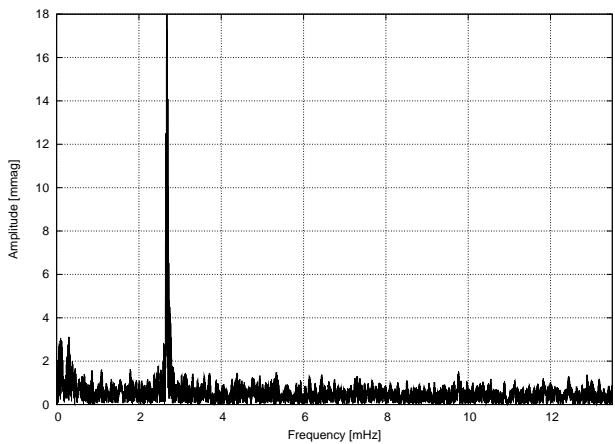}
\caption{The raw amplitude spectrum of white light data obtained for RAT0455} %% no full stop at the end
\label{fig4}
\end{figure}

\section{Newer observations and Fourier analysis}
\begin{figure}[]
%\plotone{fig5.ps}
%\includegraphics{fig5.ps}
\includegraphics[angle=-90,scale=0.3,width=\columnwidth]{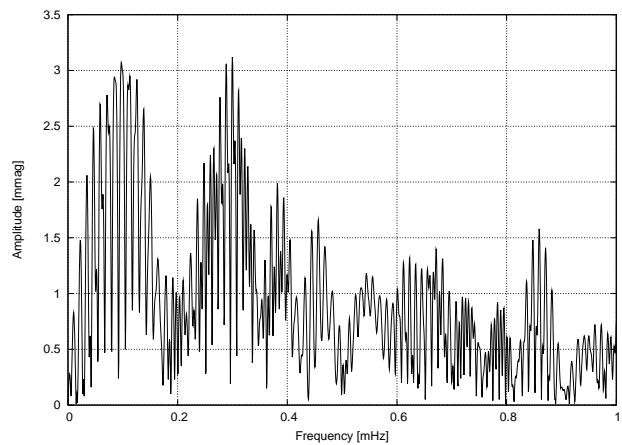}
\caption{Closer view at the low frequency region of the raw amplitude spectrum} %% no full stop at the end
\label{fig5}
\end{figure}

\begin{figure}[]
%\plotone{fig6.ps}
\includegraphics[angle=-90,scale=0.3,width=\columnwidth]{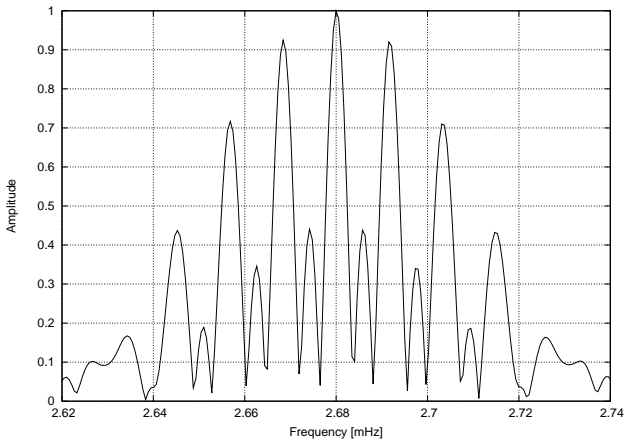}
\caption{Window function calculated on the data shifted to the frequency of the main mode} %% no full stop at the end
\label{fig6}
\end{figure}

Re\,--\,observation of the RAT0455 was performed in January 2009 using the 1.5\,m telescope located in San Pedro Martir Observatory in Mexico. We were allocated for 5 nights, however atmospheric conditions allowed us to get only 3 nights of data.
Exposure time was adjusted to 30\,sec with 37\,sec cycle time (time between two exposures). To maximize signal\,--\,to\,--\,noise ratio (S/N), no filter was used. Details of the observations are given in Tab.\ref{tab1}. The whole light curve is presented in Fig.\ref{fig2} while Fig.\ref{fig3} shows closer view at data from the last night. Variation on relatively short time scale is clearly visible. However, the light curve show some long term changes in brightness. All data were corrected for second order extinction and the amplitude spectrum was calculated. The overall amplitude spectrum, up to slightly more than the Nyquist frequency (13.5\,mHz), is shown in Fig.\ref{fig4}. As one may notice, the amplitude spectrum is dominated by one mode around 2.5\,mHz and some low-level signal at the low frequency region. Zooming in on the low frequency region is Fig.\ref{fig5}. If we recall where the p\,-- and g\,--\,modes appeared in the amplitude spectrum for other sdBV stars we can easily conclude that RAT0455 might be another hybrid star. Data taken over three nights give a frequency resolution 4\,$\mu$Hz. The window function of the data is presented in Fig.\ref{fig6}.

\begin{table*}[]
\caption{Results of the prewhitening process. For the main mode a linear change in amplitude was assumed. The amplitude for this mode is given for epoch 2454845.74943 so the phases of all modes and the derived change rate is -0.0056(5). The numbers in parentheses are the errors of the last digits.}
\begin{tabular}{cccc}
\hline
mode       & Frequency [mHz] & Amplitude [mmag] & phase [rad] \\
\hline
f$_{\rm1}$ & 2.680105(6)     & 18.6(4)          & 0.23(3) \\
f$_{\rm2}$ & 2.74959(16)     &  5.8(4)          & 4.06(7) \\
f$_{\rm3}$ & 0.29984(31)     &  3.5(4)          & 2.2(1) \\
f$_{\rm4}$ & 0.31624(41)     &  2.6(4)          & 0.7(2) \\
f$_{\rm5}$ & 0.24834(42)     &  2.2(4)          & 5.9(2) \\
\hline
\end{tabular}
\label{tab2}
\end{table*}

\begin{figure}[t]
%\plotone{fig7.ps}
\includegraphics[angle=-90,scale=0.3,width=\columnwidth]{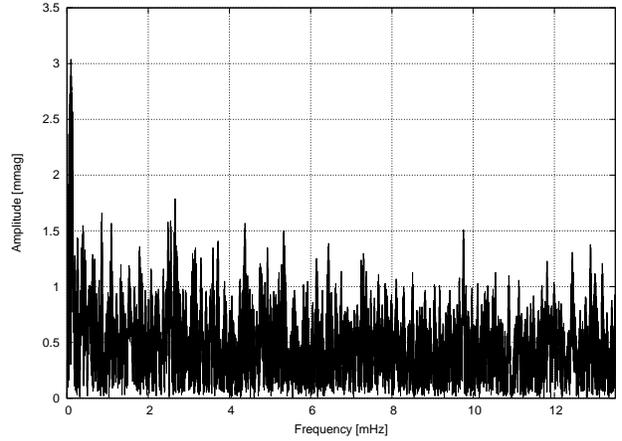}
\caption{The amplitude spectrum after removal of 5 peaks assumed to be intrinsic to the star} %% no full stop at the end
\label{fig7}
\end{figure}

Observational data were analyzed by means of Fourier analysis and prewhitening, aimed in extracting all periodicities (with S/N$>$4) which are present in the light curve. First the amplitude spectrum was calculated. Then the principal periodicity was removed from the original data. Further calculation of the amplitude spectrum was done on the data with the previous peak removed. If another peak was found, the removal step was always performed on the original data including all identified peaks. These steps were repeated until all peaks with amplitudes higher than four times the noise level were extracted. As a result of the Fourier analysis, five peaks with amplitudes, frequencies and phases (along with errors) given in Tab.\ref{tab2} were detected. In Fig.\ref{fig7} the residual amplitude spectrum after removal of all peaks from the Tab.\ref{tab2} is shown. Except one peak around 0.1\,mHz, the amplitude spectrum is dominated by the noise. The peak with 3\,mma magnitude which remained in the spectrum is most likely caused by some atmospheric effects which have not been perfectly removed from the data. A peak at this frequency is very common and has been already seen in the analysis of other pulsating stars. For this reason, it was not exctracted from the data.

\section{Discussion and conclusions}
The results presented here clearly show that RAT0455 is another pulsating hot subdwarf having hybrid nature. Only two pressure and three gravity modes have been detected. Little data and the faintness of this star have not allowed for achieving better detection threshold so there might be more modes but with small amplitudes. RAT0455 shows at least 3 modes in the g\,--\,mode region, which is more that in case of V391\,Peg but not as many as in Bal09. If more data are required to detect larger number of modes (if they really exist), that would be, together with Bal\,09, another very promising object for asteroseismology application. One has to remember, though, that to date the gravity modes in hot subdwarf hybrid stars have not been used by asteroseismology and this is still a challenge for current theoretical models.

\begin{figure}[]
\plotone{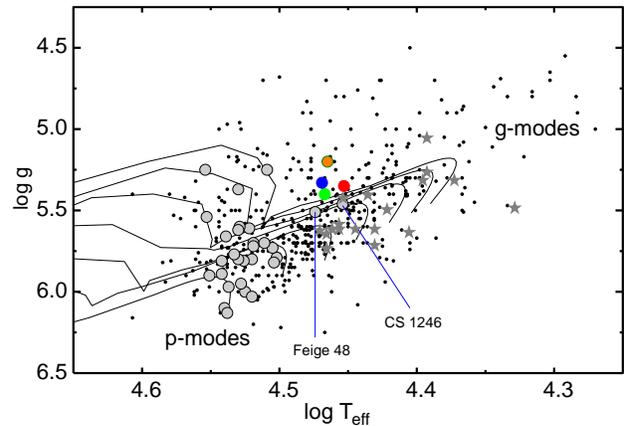}
\caption{$\log$T$_{\rm eff}$ and $\log g$ plane for sdB stars with RAT0455 assigned as hybrid star. See text for details} %% no full stop at the end
\label{fig8}
\end{figure}

If we look at the location of hybrid stars in the $\log$T$_{\rm eff}$ and $\log g$ diagram (Fig.\ref{fig8}) we can see that another object (RAT0455 marked with orange) shows hybrid nature. We can make a hypothesis that all objects in the overlapping region between p\,-- and g\,--\,modes are hybrid stars. All hybrids detected to date are plotted in Fig.\ref{fig8} and are marked with colors. In their close vicinity there are two other stars with p\,--\,modes detection only and a bunch of stars with only g\,--\,modes detected. One star from the former group is denoted as Feige\,48. This star has been observed several times but no one has reported any signal in the low frequency region which could be intrinsic to the star. However, according to S.Charpinet (private communication) newer data obtained with bigger telescopes might put new light on this issue. Detection of g\,--\,modes in this object might be more difficult as gravity modes, in general, have much smaller amplitudes compared to the pressure ones. Amplitudes of the detected modes in Feige\,48 are few mma only, so the amplitude of g modes could be to small to be found with the existing data. In addition, no one at the time this star was analyzed was expecting any signal in the low frequency domain. Even if it exists indeed, it could be removed from the data along with the long time variability usually caused by unstable atmospheric conditions. The other star, CS\,1246 has been published only recently \citep{barlow10}. Only one p\,--\,mode has been detected and the small amount of data obtained on this star has not allowed to conclude about the hybridity of this star. Discovery data revealed the dominant peak with similar period and amplitude as the discovery data of Bal\,09. Perhaps, as in case of Bal\,09, more data are needed to confirm (or discard) the hypothesis about its hybridity.

\subsection{Comparison of highest modes in high and low frequency domain}
\begin{figure}[]
\plotone{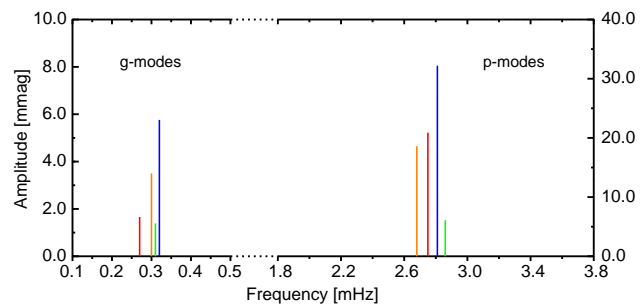}
\caption{The schematic amplitude spectrum with the dominant modes in p- and g-mode regions for all hybrid stars discovered so far. Note that vertical axis is different for both type of modes. Since the amplitudes can change from season to season by even 30\% we do not discuss differences in amplitudus. See text for more details} %% no full stop at the end
\label{fig9}
\end{figure}

If all hybrid stars are located in a well\,--\,defined region in the $\log$T$_{\rm eff}$ and $\log g$ diagram it might mean that the structure and pulsational properties could be similar. In Fig.\ref{fig9} a schematic amplitude spectrum is shown. It includes the dominant peak from the p\,-- and g\,--\,modes region. All other modes are discarded. The blue color is for Bal09, red\,--\,DW\,Lyn, green\,--\,V391\,Peg and finally the orange\,--\,RAT0455. There is a certain regularity. A pressure mode with the highest amplitude for all hybrid stars is placed almost in the same region. And the same is true for g\,--\,mode with the highest amplitude. The highest peak for CS\,1246 \citep{barlow10} is also at around 2.7\,mHz. This does not mean it is a hybrid star but does not exclude the possibility.

\subsection{Future plans}
Data taken in January 2009 revealed five modes. The noise level was about 0.5\,mmag. On the basis of research on other pulsating hot subdwarfs we know that several modes might have amplitudes below 1\,mmag level. This could mean that RAT0455 pulsates in numerous modes but the data were insufficient to show them. For this reason we have decided to re\,--\,observe this star for a longer time. In November and December 2009 we were allocated more than a week on two 2\,m class telescopes. We expect to obtain more data which allow us to detect modes with low amplitudes ($>$0.1mmag) and rotationally split multiplets if they exist. More modes can have a big impact on theoreticians who might be more encouraged to apply asteroseismology to this star.

\acknowledgments
AB appreciates financial support from the director of the Mt.\,Suhora Observatory. This work has received partial financial support from the UNAM under grant PAPIIT IN114309. Special thanks are given to the technical staff of the San Pedro M\'artir Observatory.

%\section{Citing references}

%Use \verb!\cite! command to cite reference(s).
%\smallskip

%\noindent
%\verb!\cite{baran05}! -- \cite{baran05}\\

\nocite{*}
\bibliographystyle{spr-mp-nameyear-cnd}
%\bibliography{myref}
%\bibliography{biblio-u1}

\end{document}